# Liquid marbles as thermally robust droplets: coating-assisted Leidenfrost-like effect

**Cedric Aberle, Mark Lewis, Gan Yu, Nan Lei, and Jie Xu***

*Mechanical Engineering, Washington State University, Vancouver, WA, 98686, USA. E-mail: jie.xu@wsu.edu*

**The Leidenfrost effect—prolonged evaporation of droplets on a superheated surface—happens only when the surface temperature is above a certain transitional value. Here, we show that specially engineered droplets – liquid marbles – can exhibit similar effect at any superheated temperatures (up to 465 ºC tested in our experiment) without a transition. Very possibly, this phenomenon is due to the fact that liquid marbles are droplets coated with microparticles and these microparticles help levitate the liquid core and maintain an insulation layer between the liquid and the superheated surface.**

Liquid marble is an emerging area of research in the soft matter community, as recently reviewed by McHale and Newton [1]. Liquid marbles are liquid droplets coated with self-assembled microparticles. It was first reported in 2001 by Aussillous and Quere [2], who successfully synthesized liquid marbles with lycopodium particles. Since then, liquid marbles have been fabricated using various particles, including silica, polycarbonate, polytetrafluoroethylene, polyvinylidene fluoride, polyethylene, aerogels, graphite and others [3-27].

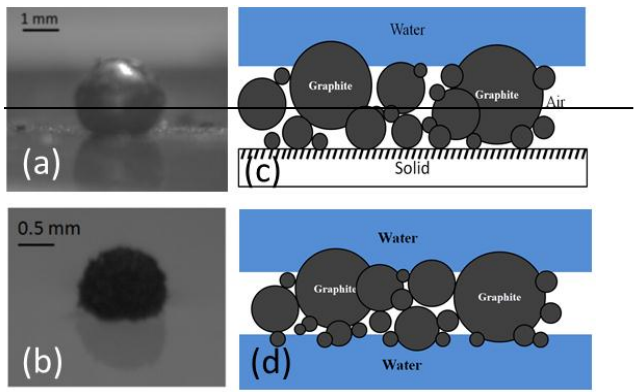

**Fig. 1** (a) a liquid marble on a solid surface; (b) a liquid marble on water surface; (c-d) illustrations of the non-wetting mechanism of liquid marbles on solid/liquid surfaces.

Thanks to the layer of coating, liquid marbles do not wet any smooth surfaces [28, 29], whether it is a solid surface (Fig. 1a) or a liquid surface (Fig. 1b), as if that surface is always superhydrophobic [30]. The schematics shown in Fig. 1c-d explain the non-wetting mechanism: the particle layer supports the weight of the droplet and the surface tension forces between the liquid and individual particles prevents the liquid from penetrating the particle layer. Note that, very possibly, the particle coating layer consists of flocs of particles instead of a monolayer, as suggested by the rough surface seen in digital camera images and indicated by previous studies [18, 27].

Droplets are widely used in many chemical and biological applications [31-35]. As specially engineered droplets, liquid marbles could be used as excellent vessels for storage and transportation of chemical reagents and bio-samples. It has been found that the manipulation of liquid marbles can be achieved by using gravitational force [28], electrostatic force [36] and magnetic force [15]. Moreover, transportation of the encapsulated liquid between liquid marbles can also be achieved as demonstrated recently by Bormashenko, *et al.*, [26], who connected two liquid marbles by a capillary tube. The flow subsequently induced by Laplace forces brings the content of one liquid marble into the other. Other interesting liquid marble applications include accelerometers [19], gas sensors [21, 22], pH sensors [16] and water pollution indicators [10].

In many chemical and biological experiments, heating is a common process. Therefore, it is very important to understand how the liquid marbles behave when they are heated. In this paper, we study the evaporation of liquid marbles when they are heated on a smooth solid surface, and compare the results to pure water droplets. Although the evaporation of liquid marbles at room temperature has been reported in Ref [8, 13], our study is the first to report the evaporation of liquid marbles when they are heated. The difference is that when evaporating at room temperature, the process is controlled by mass transfer, i.e., driven by the gradient of the vapor concentration between the liquid surface and the surroundings, and therefore the relative humidity in the surrounding is a critical parameter [37]. As for a heated evaporation, the saturation concentration at the liquid surface is much higher than the surroundings, thus the process is controlled by heat transfer and the relative humidity has negligible effects.

On the other hand, the evaporation of droplets on a heated substrate has been extensively studied. If the solid surface is at a temperature higher than the boiling point of the droplet, i.e., the surface is superheated, then the droplet will boil on the hot surface [38]. At low superheat temperatures, boiling of the droplet is so drastic that the entire droplet may vaporize completely within seconds. Contrary to intuition, at high superheat temperatures, the droplet may evaporate at a much slower rate and take much longer time to vaporize completely. This phenomenon was first reported in 1756 by J. G. Leidenfrost, a German medical doctor, and translated into English from Latin in 1966. [39] In the report, J. G. Leidenfrost studied this phenomenon by observing the boiling of water droplets on a red-hot spoon. The mechanism of the Leidenfrost phenomenon is believed to be that fast production of vapor on the side of the droplet facing the hot surface can establish a pressure field acting to repel the droplet away from the hot surface. In normal cases where the droplet is above a solid surface, the gravity of the droplet is then balanced by the repelling force from the vapor film and the droplet may hover on the vapor film. An interesting feature of Leidenfrost phenomenon is that a threshold minimum temperature is needed to have the droplet levitated. This transition temperature is termed the Leidenfrost point. In our study, we observed that the Leidenfrost-like phenomenon can take place with liquid marbles at any superheated temperatures without exhibiting a transition point.

In our study, graphite particles (Sigma Aldrich #282863, 60% of the particles are 10-20 micron sized [8]) were used for creating liquid marbles. The good thing about graphite is that it has inert reactivity, which is an important feature for handling chemical reagents. Also, the electrical conductivity and lubrication properties of graphite could be potentially useful for many lab on a chip applications. During the experiment, a precision syringe pump (KDS 210) was used to push de-ionized water at a constant flow rate (0.1 mL/min was used in our study) through a 23G needle (BD Vacutainer winged boold collection sets, Fisher Scientific) for producing water droplets through dripping. To determine the size of the droplets, an improved method based on the metrology developed in Ref [40] was used. In Ref [40], a known number of water droplets were dispensed into a container positioned on a precision balance and the mass change was recorded by the balance to deduct the weight of one droplet. However, because water in the container was exposed in the air, their results had to be corrected for evaporation. Our method is shown in Fig 2, where the same principle was used, but we covered the liquid in the container with a layer of oil, so that the evaporation can be suppressed and its effect can be ruled out.

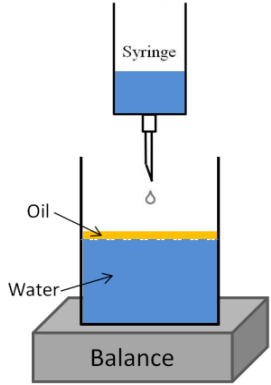

**Fig. 2** The system used to measure droplet size. A precision balance is used to measure the weight of a known number of droplets. The size of each droplet is deducted from the weight measurements. A thin layer of oil is used to increase measurement accuracy by preventing evaporation during measurement.

Table 1 lists the results of three groups of measurements performed by a precision balance (Acculab ALC). In each group, about one hundred drops were dispensed into the container that is weighed by the balance. Therefore, the weight of each droplet is obtained, and in turn, the size of each droplet is calculated. Based on this method, the average size was determined to be 2.5 mm.

**Table 1**: Droplet size measurements.

| Experiments | 1 | 2 | 3 |
|---|---|---|---|
| Number of drops | 101 | 100 | 101 |
| Total weight (g) | 0.8245 | 0.8003 | 0.8352 |
| Average weight of a single droplet (mg) | 8.16 | 8.00 | 8.27 |
| Average size of a single droplet (mm) | 2.50 | 2.48 | 2.51 |

To produce liquid marbles, water droplets generated using the aforementioned method were dripped onto a pile of graphite particles (pre-dried at 200 $^0$C overnight), and then rolled back and forth to be thoroughly coated. Once the liquid marbles were generated, they were immediately transported onto the heated substrate for experiments. The experimental system used in our study is sketched in Fig. 3, where a silicon wafer (universitywafers.com) was placed on a digital hotplate (Cimarec, Fisher Scientific) and a thermocouple (K type, Omega) taped down on the surface by a thermocouple pad (TAP, Omega) was used for monitoring the surface temperature. A digital high-speed camera (PL-B771U, Pixelink) was employed to record the evaporation process either horizontally or vertically.

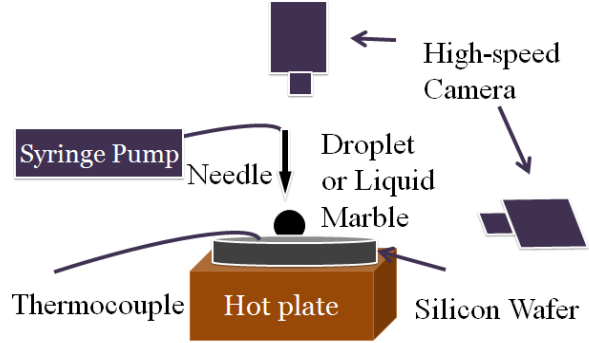

**Fig. 3** The experimental system used in our study. A silicon wafer placed on top of a precision hot plate is used to heat liquid marbles. A high speed camera is used to visualize the experiment either from top or side. A thermocouple is taped on the top surface of the wafer for monitoring the temperature.

The evaporation times of both liquid marbles and water droplets were recorded at various surface temperatures up to 465 $^o$C, which is the maximum possible temperature output from our hotplate. Fig. 4 reports the measured evaporation time of liquid marbles and water droplets. We can see that liquid marbles show great thermal robustness in all tested cases, indicated by the long evaporation time (on the order of 100 s). This is a significant improvement in comparison to water droplets, which usually evaporate within seconds, if the temperature is below the Leidenfrost point (~200 $^0$C in our case). The trend of water droplets above the Leidenfrost point closely matches that of liquid marbles. As mentioned earlier, this well-known heat transfer phenomenon is due to a vapor film that forms and levitates the water droplet away from the heated surface when the surface temperature is above the Leidenfrost point. In the Leidenfrost cases, the heat transfer is from the hot surface through a vapor layer by conduction, convection and radiation. Because the thermal conductivity of water vapor is low, this vapor layer greatly hinders heat transfer from the hot surface into the droplet. Theoretically, the total evaporation time $t_v$ for a Leidenfrost droplet can be predicted by the following equations [41]:

$$t_B^* = 1.21(V_0^*)^{5/12} \qquad (1)$$

$$V_0^* = V_0 \left(\frac{\sigma}{\rho_l g}\right)^{-3/2} \qquad (2)$$

$$t^* = t_v \left[ \frac{\rho_l^{1/2} \mu_v h_{lv}^3 \sigma^{5/2}}{k_v^3 g^{7/2} \rho_v (T_w - T_b)^3} \right]^{-1/4} \quad (3)$$

where $V_0$ is the initial droplet volume, $\sigma$ is the interfacial tension, $\rho$ is density, $g$ is the gravitational force, $\mu$ is viscosity, $h$ is enthalpy, $k$ is thermal conductivity, $T$ is temperature and the subscripts $l$, $v$, $w$ and $b$ indicate liquid, vapor, wall and boiling, respectively. These equations are obtained by solving the energy, mass, and momentum equations in the vapor film [41]. The calculated results of $t_v$ in our cases are plotted on Fig. 4, which matches our measurements for water droplets very well, although this theory cannot predict the Leidenfrost transition.

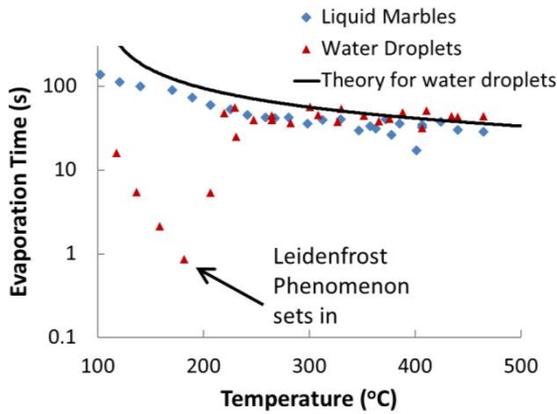

**Fig. 4** The evaporation time of liquid marbles and water droplets at different surface temperatures. Liquid marbles feature long evaporation time at all tested temperatures. Water droplets feature the same long evaporation time only when the temperature is beyond the Leidenfrost point. If the temperature is below the Leidenfrost point, water droplets usually evaporate within seconds. A theoretical curve for Leidenfrost phenomenon is also plotted for comparison.

As for liquid marbles, the precise modeling of the evaporation process seems to be difficult. This is mainly due to the experimental difficulties in both controlling the graphite coating and determining the interfacial temperature at the bottom of the liquid core.

The heat transfer between the hot surface and the liquid core is through conduction/convection/radiation in a layer of porous graphite. Because liquid marble study is still at infant stage, currently there are very limited fabrication approaches that can precisely control the structure and the thickness of the coating layer of a liquid marble [3]. Therefore, the thickness of the graphite layer in our experiments was not precisely controlled. However, we managed to roughly estimate the thickness of the layer by drying a particle-coated water film on a glass surface, and then measuring the thickness of the residues with the help of an optical microscope (Nikon MM-40), a digital camera (DXM 1200) and a measuring system (Quadra-Chek 200). The typical thickness of the graphite layer $\delta$ is then estimated to be in the range of 15-30 micron. This is on the same order of a typical Leidenfrost vapor thickness, which is in the range of 10-100 micron [42].

Another experimental difficulty is that the temperature at the interface between the liquid core and the graphite coating could not be easily measured. Temperature determination at a multiphase interface is always a difficult task [43]. Recently, an optical method based on laser thermo-reflectance has been developed to measure the temperature at the interface between liquid and solid [44]. Unfortunately, in our case, graphite particles are not transparent and therefore not suitable for any optical measurements. As a preliminary analysis, we here assume that the interfacial temperature is at the boiling point, which is normally assumed for a Leidenfrost droplet [41]. If we assume all the heat transferred into the liquid marble goes into the latent heat of vaporization (i.e., neglecting the heat loss from the marble surface due to natural convection), an energy balance requires that

$$h_{lv} \rho_l \frac{dV}{dt} = \frac{k A_b}{\delta}(T_w - T_b) \quad (4)$$

where $k$ is the conductivity of the graphite particle layer and $A_b$ is the contacting base area. This equation suggests that we can estimate $k$ by measuring the volume change rate $dV/dt$. However, unlike pure water droplets, liquid marbles undergo dramatic deformation such as buckling during evaporation (as seen from Fig. 5), which adds complexity in estimating the marble volume during the entire evaporation process. Nevertheless, we managed to measure the marble volume change before they commence buckling, which is reported in Fig. 6. This is based on the assumption that the liquid marble assumes a spherical cap shape. From Fig. 6 and using equation 4, we found that the effective thermal conductivity values of the graphite particle layer for the four temperatures (125, 175, 250 and 300 $^0$C) are 0.04, 0.10, 0.01 and 0.04 W/(mK) respectively. As expected, these values are roughly on the same order of the conductivity of water vapor, which is around 0.03 W/(mK). This simple analysis suggests that the role of particle coating is to support an insulating layer that can hinder heat transfer from the hot surface into the droplet. Note that to make it more conclusive, further experiments have to be performed. For example, experiments at even higher surface temperatures, measurements of the interfacial temperature at bottom of the liquid core and quantitative assessments of the effects from particle content in the coating layer would be desirable. That said, we believe that our report will still be enlightening to soft matter researchers, because the superior thermal robustness of liquid marbles demonstrated here, together with the innate chemical reactivity and excellent electrical conductivity of graphite, may bring up a wealth of novel applications in chemical, biological and medical studies.

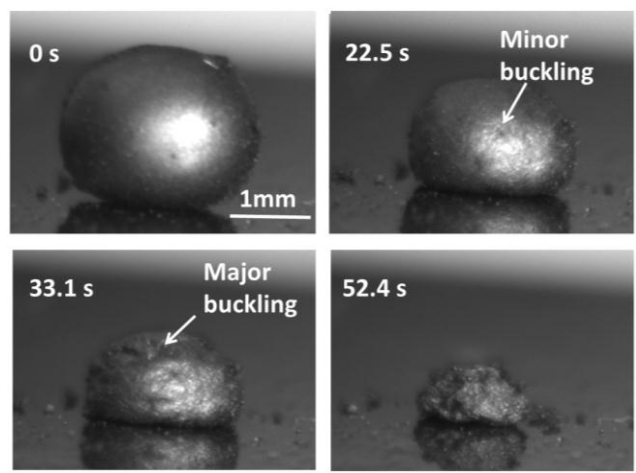

**Fig. 5** Typical deformation of a liquid marble during heated evaporation. In this case (275°C), the surface of the marble starts to buckle at about 22s and the shape of the marble undergoes dramatic deformation until it is fully dried, leaving a 3D structure of particles.

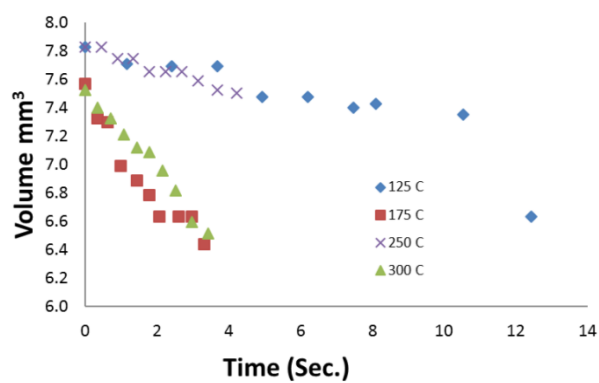

**Fig. 6** Volume variation of liquid marbles before buckling at four different temperatures, measured from high speed movies. The slopes of these curves indicate marble volume change rates, which are used to estimate the thermal conductivity of the coating layer.

On a side note, the shrinking shape of the liquid marble before buckling can be described by balancing the hydrostatic pressure and the Laplace pressure at the interface [36]. During evaporation, the direct consequence of the shrinking shape is an increase in graphite powder density on the marble surface. At some point, the surface of the liquid marble begins to collapse (as seen in the frame at 22.5 s in Fig. 5). The onset of this buckling phenomenon is probably due to the attractive Laplace force overcoming the stabilizing electrostatic forces between particles during the evaporation, as studied in a similar case of evaporating particle-filled droplets [45]. In our experiments, the onset of the buckling phenomenon seems to be random both in time and in location. This implies that the weakest point in the particle layer is randomly distributed and the strength of the weakest point is hard to predict. It is also worth noting that, since graphite is electrically conductive, the 3D structural residues could be useful for fabricating electrodes, if the evaporation process can be precisely controlled.

In summary, we investigated the heated evaporation of graphite liquid marbles on a hot substrate at various surface temperatures, and compared the results with pure water droplets. We found that if the temperature is above the Leidenfrost point, the evaporation time of liquid marbles and water droplets are almost the same, whereas if the temperature is below the Leidenfrost point, water droplets evaporate much faster and liquid marbles still exhibit the Leidenfrost-like effect. We postulate that is due to the assistance from the coating particles in maintaining a vapor layer, which insulates the droplet from the hot surface.

We thank the Undergraduate Research Mini-Grant Program (UR-MGP) at Washington State University Vancouver for financially supporting this project. We also thank Marina Reilly-Collette for proofreading this manuscript. JX thanks his elementary schoolmate Haiyan Qian who taught him to make liquid marbles from pencil leads and fountain pen ink back in 1993.